\documentclass{ifacconf}
\usepackage{graphicx}      
\usepackage{natbib}        
\usepackage{vmargin}
\setpapersize{USletter}

\setmarginsrb{0.556 in}{ 0.875 in}{ 0.875in}{0.278in}{0pt}{0pt}{.5cm}{.5cm}

\usepackage[utf8]{inputenc}
\usepackage{amsmath}
\usepackage{amsfonts}
\usepackage{amssymb}
\usepackage{float}
\usepackage{graphicx}

\usepackage[american,smartlabels,siunitx]{circuitikz}
\usepackage[american,siunitx]{circuitikz}

\usetikzlibrary{calc}
\ctikzset{bipoles/thickness=1}
\ctikzset{bipoles/length=0.8cm}
\ctikzset{bipoles/diode/height=.375}
\ctikzset{bipoles/diode/width=.3}
\ctikzset{tripoles/thyristor/height=.8}
\ctikzset{tripoles/thyristor/width=1}
\ctikzset{bipoles/vsourceam/height/.initial=.7}
\ctikzset{bipoles/vsourceam/width/.initial=.7}
\tikzstyle{every node}=[font=\small]
\tikzstyle{every path}=[line width=0.8pt,line cap=round,line join=round]

\def\hal{{1 \over 2}}

\def\begequarr{\begin{eqnarray}}
\def\endequarr{\end{eqnarray}}
\def\begequarrs{\begin{eqnarray*}}
\def\endequarrs{\end{eqnarray*}}
\def\begarr{\begin{array}}
\def\endarr{\end{array}}
\def\begequ{\begin{equation}}
\def\endequ{\end{equation}}
\def\lab{\label}
\def\begdes{\begin{description}}
\def\enddes{\end{description}}
\def\begenu{\begin{enumerate}}
\def\begite{\begin{itemize}}
\def\endite{\end{itemize}}
\def\endenu{\end{enumerate}}

\def\l2{{\mathcal L}_2}
\def\l2e{{\cal L}_{2e}}

\def\rea{\mathbb{R}}

\def\lef[{\left[\begin{array}}
\def\rig]{\end{array}\right]}

\usepackage{color}

\newcommand{\ju}[1]{{\color{black} #1}}
\newcommand{\jo}[1]{{\color{black} #1}}

\usepackage[prependcaption,colorinlistoftodos]{todonotes}

\begin{document}

\begin{frontmatter}
%
%
\title{Active Damping of a DC Network with a Constant Power Load: An Adaptive Passivity-based Control Approach\thanksref{footnoteinfo}}
\thanks[footnoteinfo]{The work of Juan E. Machado was financially supported by the {\em National Council of Science and Technology} (CONACyT for its acronym in Spanish) from Mexico. The work of Jos{\'e}~Arocas-P{\'e}rez and Robert Gri\~n\'o was partially supported by the Government of Spain through the \emph{Agencia Estatal de Investigaci{\'o}n} Project DPI2017-85404-P and by the \emph{Generalitat de Catalunya} through the Project 2017 SGR 872.
}
\thanks[footnoteinfo]{\ju{An abridged version of this paper has been submitted to Congreso Nacional de Control Automatico 2018, 10--12 October 2018, San Luis Potos\'i, M\'exico.}}

\author[L2S]{Juan E. Machado} 
\author[IOC]{Jos{\'e}~Arocas-P{\'e}rez} 
\author[SU]{Wei He}
\author[L2S]{Romeo Ortega}
\author[IOC]{Robert~Gri{\~n}{\'o}}

\address[L2S]{Laboratoire des Signaux et Syst\`emes, CNRS-Sup\'elec, Plateau du Moulon, 91192, France (e-mail: juan.machado@l2s.centralesupelec.fr, romeo.ortega@lss.supelec.fr).}
\address[IOC]{Institute of Industrial and Control Engineering (IOC), Universitat Polit\`{e}cnica de Catalunya, 08028 Barcelona, Spain (e-mail: jose.arocas@upc.edu, roberto.grino@upc.edu).}
\address[SU]{Key Laboratory of Measurements and Control of Complex Systems of Engineering, Ministry of Education, School of Automation, Southeast University, 210096 Nanjing, China (e-mail: hwei@seu.edu.cn)}
%
%
\begin{abstract}
	This paper proposes a nonlinear, adaptive controller to increase the stability margin of a direct-current (DC) small-scale electrical network containing a constant power load, whose value is unknown. Due to their negative incremental impedance, constant power loads are known to reduce the effective damping of a network, leading to voltage oscillations and even to network collapse. To tackle this problem, we consider the incorporation of a controlled DC-DC power converter between the feeder and the constant power load. The design of the control law for the converter is based on the use of standard Passivity-Based Control and Immersion and Invariance theories. The good performance of the controller is evaluated with numerical simulations.
\end{abstract}
\end{frontmatter}

\section{Introduction}
This note is concerned with the stability analysis of electrical networks with Constant Power Loads (CPLs). It is well-known that, due to their negative impedance characteristic, CPLs induce voltage oscillations or even network collapse, see \citep{Emadi2006}. The analysis of networks with these type of loads started with the work of \citep{Middlebrook} and has been an active research problem since then, {\em e.g.},  \citep{Belkhayat1995}, \citep{Belkhayat1995a} and \citep{Emadi2006}. We refer the reader to \citep{singh2017} and the references therein for a recent review on this topic.

The stability analysis of networks with CPLs has been carried out using different approaches. Linearization methods were used in \citep{Anand2013}, \citep{BARetal}, and \citep{Marxetal2012}, see also \citep{arocas2017local} . Nonlinear techniques such as the Brayton-Moser mixed potential theory, introduced in \citep{Bryton1964}, has been used in  \citep{Belkhayat1995} and \citep{Cavanagh2017} to derive sufficient conditions for stability. The use of this technique, to estimate regions of attraction, is reviewed in  \citep{Marxetal2012}. More recently,  in \citep{pooyaetal-2018}, using the framework of port-Hamiltonian (pH) systems, see  \citep{vanderSchaft2017}, sufficient conditions for stability are presented. It is shown that by imposing upper bounds on the CPLs maximum power, the stability analysis can be concluded using the shifted Hamiltonian as a candidate Lyapunov function.  This approach was firstly explored in \citep{Bayu2007} for general nonlinear systems.

Various controller design techniques have been proposed to enlarge the domain of attraction of this kind of networks with CPLs---a review may be found in \citep{singh2017}. These stabilization techniques can be divided into passive and active {\em damping} methods. The former is based on open-loop hardware modifications, whereas the latter implies the modification of existing or added control loops, which may imply the interconnection of additional hardware. In this note we follow the ideas presented in \citep{carmeli-2012}  and \citep{Zhangetal2013}. In these works, a connection of a {\em controlled} power converter, in parallel with the CPL (shunt damper), is proposed to increase the stability margins of a small-scale DC network.  In \citep{carmeli-2012}, assuming a simplified model for the converter, idealized as a controlled current source, a linear control law is designed. In  \citep{Zhangetal2013}, a full model for the power converter is used. Nonetheless, their stabilization result is based on linearization. A large signal stability analysis, but using approximate techniques, such as the Takagi-Sugeno fuzzy model, is carried out in \citep{kim-2016}.

The main contribution of the present note is to propose a physically realizable nonlinear control law that renders a small-scale DC network, containing a CPL, stable for a wide range of  power consumption values from this load. The construction of the controller relies on the use of standard Pasivity-Based Control (s-PBC) theory  \citep{ortega2013passivity}. Additionally, using Immersion and Invariance (I\&I) theory \citep{astolfi2007nonlinear} the controller is made {\em adaptive} by including an estimator of the power consumed by the CPL, which is generally {\em unknown} in a practical context. The good performance of the proposed adaptive controller is evaluated with realistic numerical simulations.

The rest of the paper is structured as follows. In Section \ref{sec: problem description} we give the model of a small-scale network containing a CPL, analyze its equilibria and describe the stability problem addressed in the paper. The proposed controller configuration, adopted from  \citep{carmeli-2012}  and \citep{Zhangetal2013}, is presented in Section \ref{sec: objectives and methodology}. Our main stabilization results are included in Section \ref{sec: main results}. In Section \ref{sec: numerical validation} we present the numerical implementation of our theoretical developments. The note is finalized in Section \ref{sec: conclusions and future work} with a number of concluding remarks and open problems that can be addressed to extend our results.
%
\section{Problem Formulation}\label{sec: problem description}
\lab{sec2}
%
\subsection{Description of the \jo{open-loop} system}
\begin{figure}[H] 	 	
\begin{center}
\includegraphics[width=.66\linewidth]{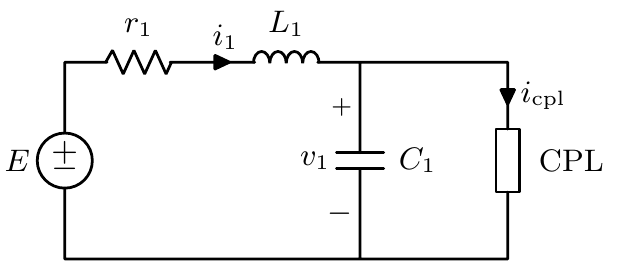}
\end{center}
	\caption{\normalsize A DC source \jo{supplying power to} an instantaneous CPL.}
	\label{fig: dc network with cpl}
\end{figure}

\lab{subsec21}
%
We study a simplified model of a DC power system as shown in Fig. \ref{fig: dc network with cpl}. This simple model has been used in the literature, {\em e.g.}, in \citep{Zhangetal2013}, \citep{mosskul-2015} and  \citep{wu-2015}, to study the stability problems associated with CPLs. It consists of a DC voltage source supplying electric energy to an instantaneous CPL. The transmission line is simply represented by a lossy inductor and the CPL is connected through a bus capacitor. The dynamic model for this network is given by
\begin{equation}\label{eq: net dyn i-v coord}
\begin{aligned}
L_1\dot{i}_1 & =-r_1i_1-v_1+E,\\
C_1\dot{v}_1 & = i_1-\frac{P}{v_1},
\end{aligned}
\end{equation}
where $i_1$ and $v_1$ denote the current of the inductor $L_1>0$ and the voltage of the capacitor $C_1>0$, respectively. \jo{The {\em constant} parameter $P$ corresponds to the power extracted or injected  into the network by the CPL, being positive in the former case and negative in the latter. In the sequel, we focus our attention in the critical case $P\geq0$.}

The state space for this system is defined as follows
\begin{equation*}
\mathcal{X}_1:=\{(i_1,v_1)\in\mathbb{R}^{2}:~ v_1>0 \}.
\end{equation*}

\subsection{Equilibrium analysis \jo{of the open-loop system}}
\lab{subsec22}
%
The following proposition pertains to the existence of steady states for the system \eqref{eq: net dyn i-v coord} whose proof, being straightforward, is omitted for brevity.

\begin{prop}\label{prop: equlibria}
The system \eqref{eq: net dyn i-v coord} admits two equilibrium points, which are given by
\begin{equation}
\lab{equpoi}
  \bar i_1 =\frac{E\mp\sqrt{\Delta}}{2r_1},~ \bar v_1=\frac{E\pm\sqrt{\Delta}}{2},
  \end{equation}
Furthermore, these equilibrium points are real if and only if
\begin{align}\label{eq: exis steady states power}
\Delta:=E^2-4Pr_1\geq 0~ \Leftrightarrow~ P \leq \frac{E^2}{4r_1}.
\end{align}
\end{prop}

\begin{rem}
The system has two equilibria. However, we are mainly interested in operating the system in the equilibrium  with the  highest value for the voltage $\bar v_1$ and the lowest value for the current $\bar i_1$. \jo{In the sequel,} whenever we write $(\bar i_1, \bar v_1)$, we are making reference to this particular equilibrium point. In the next section, \jo{a controller is added to the system and} the control objectives are defined, we will propose a desired value for $\bar v_1$ to be stabilized.
\end{rem}

In the next proposition we give sufficient and necessary conditions for the equilibrium point $(\bar i_1,\bar v_1)$ to be stable. \jo{This result follows directly from studying the eigenvalues of the system \eqref{eq: net dyn i-v coord} at $(\bar i_1,\bar v_1)$.}. 

\ju{
	\begin{prop}\label{prop: stab one port open loop}
		For the system \eqref{eq: net dyn i-v coord}, assume that $C_1< \frac{L_1}{r_1^2}$, then, a necessary condition for  $(\bar i_1,\bar v_1)$ to be stable is given by
		\begin{equation}\label{eq: P upper bound - necessary no damper}
			P\leq \frac{{E}^2 C_1 L_1 r_1}{\left(L_1+C_1 r_1^2\right){}^2}.
		\end{equation}
		Furthermore, if this inequality holds strictly, then, $(\bar i_1,\bar v_1)$ is also asymptotically stable. Lastly, in the case that $C_1\geq  \frac{L_1}{r_1^2}$, the condition
		$$
		P\leq\frac{E^2}{4r_1},
		$$
		is necessary and sufficient for $(\bar i_1,\bar v_1)$ to be stable.
\end{prop}}
\subsection{Control objectives}
\lab{subsec23}
%
To streamline the presentation of our control objectives we make the following observations. 

\begite
\item[(i)] As seen in \eqref{equpoi} \ju{and \eqref{eq: exis steady states power}}, the equilibrium points depend on the value of the parameter $P$. In particular, the value of $\bar v_1$ decreases when $P$ increases.

\item[(ii)] Proposition \ref{prop: stab one port open loop} shows that, when the capacitance $C_1$ is not big enough, \ju{then, to maintain stability, the power extraction from the CPL must be strictly smaller than the upper bound for existence of equilibria given in \eqref{eq: exis steady states power}.}

\endite
 
In the light of these remarks our control objectives are specified as follows.

\begin{itemize}
\item[CO1] \ju{Stabilize the voltage $v_1$ around a desired value.}
\item[CO2] Relax the upper bound for $P$ established in \eqref{eq: P upper bound - necessary no damper}.
\item[CO3] Achieve these objectives without the knowledge of $P$.
\end{itemize}

Condition (b) of Proposition \ref{prop: stab one port open loop} suggests a passive method to achieve these objectives, which consists in increasing the effective capacitance $C_1$. This can be done with the connection in parallel of a suitable capacitor and the CPL. Some disadvantages of this approach are mentioned in \cite[Section III.A]{carmeli-2012}. Instead of this passive approach, we propose to add a {\em switched capacitor}. More precisely, following  \citep{carmeli-2012}  and \citep{Zhangetal2013}, we add a shunt damper, that is a power converter in parallel with the CPL. The derivation of the control law for this converter, that ensures the control objectives above, is the main contribution of the paper.

%
\section{Augmented Circuit Model}
\label{sec: objectives and methodology}
\begin{figure}[H] 	 	
\begin{center}
\includegraphics[width=0.9\linewidth]{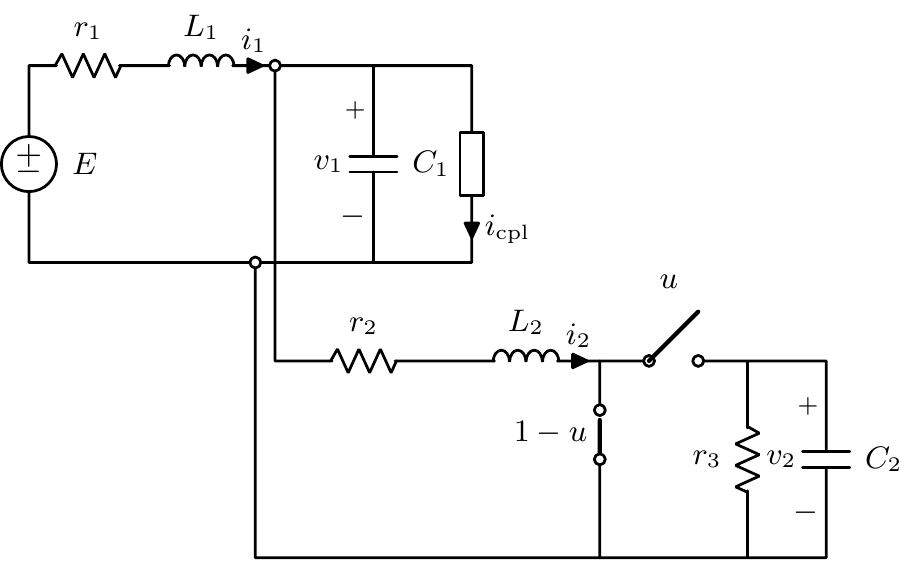}
\end{center}
	\caption{\normalsize A shunt damper connected between the feeder and the load for the network of Fig. \ref{fig: dc network with cpl}. \jo{For the switches,  ``1" means closed and ``0" means open.}}
	\label{fig: shunt damper}
\end{figure}

%
As  proposed in \citep{carmeli-2012} and \citep{Zhangetal2013}, we consider the addition of a controlled shunt damper between the feeder and the load, as shown in Fig. \ref{fig: shunt damper}. The damper consists of a DC-DC power converter, composed of two complementary switches $u$ and $(1-u)$, a lossy inductor $L_2$ and a capacitor $C_2$. The losses associated with the switching devices are modeled with the resistor $r_3$.

The averaged dynamic model of the interconnected system shown in Fig. \ref{fig: shunt damper} is given by
\begin{equation}\label{eq: dyn net + SD}
\begin{aligned}
L_1\dot{i}_1 & =-r_1i_1-v_1+E,\\
C_1\dot{v}_1 & = i_1-\frac{P}{v_1}-i_2,\\
L_2\dot{i}_2 & = -r_2i_2-uv_2+v_1,\\
C_2\dot{v}_2 & = -\left(\tfrac{1}{r_3}\right)v_2+ui_2,
\end{aligned}
\end{equation}
where $i_2$ is the current of the inductor $L_2>0$, $v_2$ is the voltage of the capacitor $C_2>0$, and the variable $u\in[0,1]$ represents the duty cycle, which is the control variable for the system.\

Before closing this section, and for ease of reference in the sequel, we notice that the system \eqref{eq: dyn net + SD} may be written in the classical form $\dot x=f(x)+g(x)u$ where
\begin{equation*}
x:=\text{col}(i_1,v_1,i_2,v_2),
\end{equation*}
is the state vector and
\begin{align*}
f(x) & := {{D}^{-1}}\begin{bmatrix}
	 -r_1x_1-x_1+E\\
	 x_1-\frac{P}{x_2}-x_3\\
	 -r_2x_3+x_2\\
	 -\frac{1}{r_3}x_4
	 \end{bmatrix},\;
g(x)  := {{D}^{-1}}\begin{bmatrix}
0\\
0\\
-x_4\\
{x_3}
\end{bmatrix},
\end{align*}
{with $D:=\text{diag}\{L_1,C_1,L_2,C_2 \}$ and state space
\begin{align*}
\mathcal{X} & :=\{x\in\mathbb{R}^{4}:~x_2>0,~ x_4>0 \}.
\end{align*}
%
\section{Main results}
\label{sec: main results}
%
In this section we propose a nonlinear adaptive controller that ensures that the network under study satisfies the control objectives of Subsection \ref{subsec23}. Towards this end, we first analyze the set of assignable equilibria and establish constraints on the system parameters for the existence of physically realizable steady states. Second, following s-PBC theory, we design a  control law that asymptotically stabilizes a desired equilibrium state assuming the CPL power $P$ is known. Finally, using I\&I theory, we present an estimator for $P$ that makes adaptive the proposed controller, preserving the stability property.
\subsection{Assignable equilibria}
\label{sec: equilibria analysis with shunt damper}
%
We say that a pair\footnote{Following a standard procedure, we let $u$ live in $\rea$ even though, being a duty cycle, it is restricted to the set $[0,1]$. This issue is partially addressed in Corollary \ref{cor: choice of bar x2}.} $(\bar{x},\bar{u})\in\mathcal{X}\times \rea$ is an equilibrium  of \eqref{eq: dyn net + SD} if and only if
\begin{equation}\label{eq: equilibrium equation ss}
f(\bar{x})+g(\bar{x})\bar{u} = 0.
\end{equation}
We define the set of {\em assignable equilibria} $\mathcal{E}$, as the set of points $\bar{x}\in \mathcal{X}$ for which there exists $\bar{u}\in \rea$ such that \eqref{eq: equilibrium equation ss} holds. This set can be computed as follows (see \cite[Lemma 2]{cbi-ortegaetal-2008}). Let $g^\perp:\mathcal{X}\rightarrow \mathbb{R}^{3\times 4}$ be a full-rank left-annihilator of $g$, {\em i.e.}, it satisfies $g^\perp(x)g(x)=0$ for all $x\in\mathcal{X}$ \jo{in its rank}. Then, $\mathcal{E}$ is given by
\begin{equation}\label{eq: set assign equil}
\mathcal{E}=\{{x}\in\mathcal{X}:~ g^\perp({x})f({x})=0\}.
\end{equation}
Furthermore, the associated unique equilibrium input $\bar{u}$ is given by
\begin{equation}\label{eq: equil input value}
\bar{u}=-\left(g^\top(\bar{x})g(\bar{x}) \right)^{-1}g^\top(\bar{x})f(\bar{x}).
\end{equation}

In the following proposition we derive the explicit values of $\bar{x}\in\mathcal{E}$ and its associated $\bar{u}$, which are compatible with the control objectives.

\begin{prop}\label{prop: assign equil}
Fix $\bar{x}_2>0$ as a desired operation value for the network. Then, $\bar{x}\in\mathcal{E}$ if and only if
\begin{equation}\label{eq: bounds for P in terms of x2bar}
P_M(\bar{x}_2) - \frac{\bar{x}^2_2}{r_2}<P<P_M(\bar{x}_2),
\end{equation}
where
\begin{equation*}
P_M(\bar{x}_2):=\frac{\bar{x}_2}{r_1}(E-\bar{x}_2).
\end{equation*}
In that case, we \jo{can parametrize the remaining equilibrium components as}
\begin{equation}\label{eq: bar x assign equi}
\begin{aligned}
\bar{x}_1 &=\frac{E-\bar x_2}{r_1},\\
\bar{x}_3 &= -\frac{Pr_1-E\bar{x}_2+\bar{x}_2^2}{r_1\bar x_2},\\
\bar{x}_4 & =\frac{1}{r_1\bar{x}_2} \sqrt{r_3\kappa_1(\bar{x}_2,P)\kappa_2(\bar{x}_2,P)},
\end{aligned}
\end{equation}
where
\begin{align*}
\kappa_1(\bar{x}_2,P) & :=\jo{-\bar{x}_2^2+E\bar{x}_2-r_1P,}\\
\kappa_2(\bar{x}_2,P) & :=\jo{(r_1+r_2)\bar{x}_2^2-r_2E\bar{x}_2+r_1r_2P.}
\end{align*}
Furthermore, the associated equilibrium value for the input variable is given by
\begin{equation*}
\bar{u}=\sqrt{\frac{\kappa_2(\bar{x}_2,P)}{r_3\kappa_1(\bar{x}_2,P)}}.
\end{equation*}
\end{prop}
\begin{pf}
The proof follows straightforward if we define
\begin{equation*}
g^\perp(x)=\left[
\begin{array}{cccc}
 1 & 0 & 0 & 0 \\
 0 & 1 & 0 & 0 \\
 0 & 0 & x_3& x_4
\end{array}
\right],
\end{equation*}
and we compute $\mathcal{E}$ and $\bar{u}$ from \eqref{eq: set assign equil} and \eqref{eq: equil input value}, respectively. \qed
\end{pf}

\subsection{Equilibrium for maximum power extraction and control realizability}
\label{subsec42}
%
Observe from \eqref{eq: bounds for P in terms of x2bar} that the amount of power that can be extracted by the CPL is limited by the choice of $\bar{x}_2$. Analogously, the equilibrium value for $u$, given in  \eqref{eq: equil input value}, depends both on $\bar{x}_2$ and on $P$. In the following corollary we choose a  desired value of $\bar{x}_2$ which permits a maximum extraction of power from the CPL. Additionally, we establish conditions on $P$ which guarantee that $\bar{u}$ is strictly smaller than one, {\em i.e.}, physically realizable with the power converter. The proof of the corollary, being straightforward, is omitted for brevity.

\begin{cor}\label{cor: choice of bar x2}
The largest admissible extracted power $P_M(\bar{x}_2)$ is maximized with the choice  
\begin{equation}\label{eq: desired bar x2 optimal P}
\bar{x}_2=\frac{E}{2}.
\end{equation}
Given this value, $\bar{x} \in \mathcal{E}$ if and only if the extracted power satisfies
\begin{equation*}
\frac{E^2(r_2-r_1)}{4r_1r_2}<P<\frac{E^2}{4r_1}.
\end{equation*}
Furthermore, the associated equilibrium value for $u$, given by
\begin{equation*}
\bar{u}=\sqrt{\frac{{E}^2 ({r_1}-{r_2})+4 P {r_1} {r_2}}{r_3 \left({E}^2-4 P {r_1}\right)}},
\end{equation*}
is {\em strictly smaller than one} if and only if the upper-bound on $P$ is restricted even further to
\begin{equation}\label{eq: new upper bound for P}
P<\frac{E^2(r_2+r_3-r_1)}{4r_1(r_2+r_3)}.
\end{equation}
\end{cor}
\subsection{Design of a stabilizing control law}
\lab{subsec43}
In this subsection we present a control law that renders \jo{the desired equilibrium point \eqref{eq: bar x assign equi}, with $\bar{x}_2$ given in \eqref{eq: desired bar x2 optimal P}, asymptotically stable.} The controller design is carried out following the s-PBC methodology.\

{
For a better readability and following the ideas presented in \citep[Section IV]{cisnerosetal-2013}, we show, \jo{under the assumption that $x_2(t),x_4(t)>0$ for all $t$}, that there exists a suitable change of variables for $u$, which \ju{allows us to write the system} \eqref{eq: dyn net + SD} in the cascade form shown in Fig \ref{fig: cascaded_system}.
\begin{figure}[H] 	 	
	\begin{center}
		\includegraphics[width=0.5\linewidth]{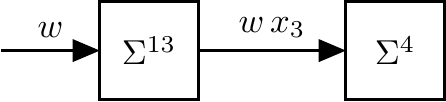}
	\end{center}
	\caption{\normalsize Block diagram for the cascaded interconnection between the subsystems \eqref{eq: sub 13} and \eqref{eq: sub 4}.}
	\label{fig: cascaded_system}
\end{figure}
\begin{prop}\label{prop: cascade decomp}
Define the auxiliary control variable
\begin{equation*}
w=x_4u.
\end{equation*}
Then, in the state space $\mathcal{X}$, the system \eqref{eq: dyn net + SD}  admits cascade decomposition into the subsystems
\begin{equation}\label{eq: sub 13}
\jo{\Sigma^{13}}  :\left\{\begin{matrix}
L_1\dot{x}_1 & = &-r_1x_1-x_2+E,\\
C_1\dot{x}_2 & = & x_1-\frac{P}{x_2}-x_3,\\
L_2\dot{x}_3 & = & -r_2x_3+x_2-w,
\end{matrix} \right.
\end{equation}
and
\begin{equation}\label{eq: sub 4}
\jo{\Sigma^{4}}  : \begin{matrix}
C_2\dot{x}_4 & = & -(\tfrac{1}{r_3})x_4+\frac{w}{x_4}x_3.
\end{matrix}
\end{equation}
\end{prop}

Now, we make the important observation that the system $\jo{\Sigma^{13}}$ admits an Euler-Lagrange representation. For more details on the representation of electrical circuits in this formalism, we refer the interested reader to \citep{ortega2013passivity}.
\begin{prop}
Define the vector
\begin{equation*}
\jo{x^{13}}:=\text{col}(x_1,x_2,x_3),
\end{equation*}
the matrices
\begin{equation*}
\mathcal{D}:=\text{diag}\left\{{L_1},{C_1},{L_2} \right\}>0,
\end{equation*}
\begin{equation*}
\mathcal{C}:=\begin{bmatrix}
0 & 1 & 0 \\
-1 & 0 & 1 \\
0 & -1 & 0 \\
\end{bmatrix},~~~ \jo{R(x^{13})}:=\begin{bmatrix}
r_1 & 0 & 0 \\
0 & \frac{P}{x_2^2} & 0 \\
0 & 0 & r_2 \\
\end{bmatrix},
\end{equation*}
and the constant vectors
\begin{align*}
 \mathcal{K}:=\text{col}(E,0,0),
 \mathcal{G}=\text{col}(0,0,-1).
\end{align*}
Then, the system $\jo{\Sigma^{13}}$ admits an Euler-Lagrange formulation given by
	\begin{equation*}
    \begin{aligned}
    \mathcal{D}\dot x^{13} +(\mathcal{C}+R(x^{13}))x^{13}=\mathcal{K}+\mathcal{G}w.
    \end{aligned}
    \end{equation*}
\end{prop}
The importance of this representation is that it allows a direct application of the s-PBC methodology  to define a control law for $w$ which renders the equilibrium point $\jo{\bar{x}^{13}}$ of $\jo{\Sigma^{13}}$ exponentially stable. Consider the following

\begin{prop}\label{prop: spbc subsystem 1}
For the system $\jo{\Sigma^{13}}$, assign the static state-feedback control
\begin{equation*}
w=\phi_2(\jo{x^{13}}),
\end{equation*}
where\jo{
	\begin{align*}
	\phi_1(x_2) & := \bar{x}_1-\tfrac{P}{x_2^2}\bar{x}_2+ k_1(x_2-\bar{x}_2),\\
	\phi_2^\text{a} & :=  -r_2{\phi_1(x_2)}-L_2\left(k_1+2\tfrac{P\bar{x}_2}{x_2^3}\right)f_2(x_1,x_2,x_3), \nonumber\\
	\phi_2^\text{b} & :={\bar{x}_2}+{k_2}\left(x_3-\phi_1(x_2)\right), \nonumber \\
	\phi_2(x^{13}) & := \phi_2^\text{a}+\phi_2^\text{b},
	\end{align*}}
and $k_1,k_2\geq 0$ are arbitrary constants. Then, the equilibrium $\jo{\bar x^{13}}$ is exponentially stable.
\end{prop}

\begin{pf}
Following an s-PBC methodology, define
\begin{equation*}
\hat{x}:=(\bar{x}_1,\bar{x}_2,\phi_1(x_2)),
\end{equation*}
and the added damping matrix
\begin{equation*}
\mathcal{R}_\text{a}:=\text{diag}(0,k_1,k_2)\geq 0.
\end{equation*}
Fix $w=\jo{\phi_2(x^{13})}$. Then, $\hat{x}$ satisfies the expression
\begin{equation*}
\mathcal{D}\dot{\hat{x}}+\left(\mathcal{C}+R(\jo{x^{13}})\right)\hat{x}+\mathcal{R}_\text{a}\left(\hat{x}-\jo{x^{13}}\right)=\mathcal{K}+\mathcal{G}\cdot\phi_2(\jo{x^{13}}),
\end{equation*}
Define $e:=\jo{x^{13}}-\hat{x}$. Its derivative satisfies
\begin{equation}\label{eq: error dynamics}
\mathcal{D}\dot{e}+\left(\mathcal{C}+\mathcal{R}_\text{d}(\jo{x^{13}}) \right)e=0.
\end{equation}
where we defined the desired damping matrix
\begin{align*}
\mathcal{R}_\text{d}(\jo{x^{13}}) & := R(\jo{x^{13}})+\mathcal{R}_\text{a} \nonumber \\
 & = \text{diag}\{r_1,\frac{P}{x_2^2}+k_1,r_2+k_2 \},
\end{align*}
which is {\em positive definite} for all $x_2>0$. Consider the positive definite function
$$
V(e):=\hal e^\top \mathcal{D}e.
$$
Its derivative, along solutions of \eqref{eq: error dynamics}, satisfies
$$
\dot V= -e^\top \mathcal{R}_\text{d}(x)e \leq - c V,
$$
for a constant $c>0$, provided $x_2>0$. From the last inequality we immediately conclude that the origin of \eqref{eq: error dynamics} is exponentially stable \cite[Theorem 4.10]{khalil3ed}. Using the fact that $\phi_1(\bar x_2)=\bar x_3$ we conclude that $\jo{\bar x^{13}}$ is an exponentially stable equilibrium point of $\jo{\Sigma^{13}}$. \qed
\end{pf}

%
%
%

The next step in our stabilization result, is to show that, when $w=\phi_2(\jo{x^{13}})$ , then $x_4$, solutions of the subsystem $\Sigma^4$, converge exponentially to $\bar x_4$.\

\begin{prop}\label{prop: stab subs sigma 4}
	Let \jo{$x^{13}(t)$ be any solution} of the subsystem  $\jo{\Sigma^{13}}$ in closed-loop with state-feedback $w=\phi_2(\jo{x^{13}})$. Then, \jo{every solution $x_4(t)$ of $\Sigma^{4}$, converges} exponentially to $\bar x_4$.
\end{prop}
\begin{pf}
Consider the change of coordinates	
\begin{equation*}
z=\frac{1}{2}C_2x_4^2.
\end{equation*}
Then, 
\begin{equation}
\lab{dotz}
\dot{z}=-\frac{2}{r_3C_2}z+\phi_2(\jo{x^{13}})x_3.
\end{equation}
From Prop. \ref{prop: spbc subsystem 1}, we have that $\jo{\bar x^{13}}$ is an exponentially stable equilibrium point of $\jo{\Sigma^{13}}$. This implies that the term $\phi_2(\jo{x^{13}})x_3$ remains bounded and converges, exponentially, to the value 
\begin{align*}
\phi_2(\jo{\bar x^{13}}) \bar x_3 & =\left(-r_2\bar x_3+\bar x_2 \right)\bar x_3\\
& = \left(\bar u \bar x_4 \right)\bar x_{3},
\end{align*}
where we have used the steady state expression
\begin{align*}
 -r_2{\bar{x}_3}-\bar{u}\bar{x}_4+\bar{x}_2=0.
\end{align*}
It follows trivially from \eqref{dotz} that
\begin{equation*}
z\rightarrow  \frac{r_3C_2}{2}\phi_2(\jo{\bar{x}^{13}})\bar x_3=\frac{r_3C_2}{2}\bar{u}\bar x_4\bar x_3.
\end{equation*}
exponentially. Using the steady state expression
$$\bar{u}\bar{x}_3-\frac{1}{r_3}\bar{x}_4=0,$$
and restricting $x_4>0$, we conclude that $x_4\rightarrow \bar x_4$ exponentially. \qed
\end{pf}
As a direct application of Props. \ref{prop: spbc subsystem 1} and \ref{prop: stab subs sigma 4}, we obtain our main result.

\begin{prop}\label{prop: stabilization result ideal P}
Consider the system \eqref{eq: dyn net + SD}  in closed-loop with the static state-feedback control
\begin{equation}\label{eq: u simpler }
u=\gamma(x),
\end{equation}
where
\begin{align}
\gamma(x) & = \tfrac{1}{ x_4 }\phi_2(x_1,x_2,x_3). \label{eq: gamma}
\end{align}
Then, $\bar{x}\in\mathcal{E}$ is a locally, exponentially stable equilibrium point of the closed-loop system.\
\end{prop}
}

\subsection{Stabilization assuming an unknown CPL power}\label{sec: power estimator and stab analysis}

In this section we propose  an adaptive version of the previously designed controller.  Now, we assume that the CPL power $P$ is constant but {\em unknown}. First, following an I\&I technique, a dynamic estimate for $P$, which we denote \jo{by $\hat{P}(t)$}, is presented. We show that the error between the estimate and the actual value of $P$ converges to zero exponentially fast for any initial condition.

\begin{prop}\label{prop: power estimator}
For the system \eqref{eq: dyn net + SD}, assume that $P$ is unknown. Define an on-line estimate for $P$ as follows
\begin{align}\label{eq: estimate for P}
\hat{P}(t) & =-\frac{1}{2}\cdot\jo{k_3} C_1v_1^2+P_\text{I}(t),\\
\dot{P}_\text{I}(t) & = \jo{k_3} v_1\left(i_1-i_2 \right)+\frac{1}{2}{\jo{k_3}^2}C_1v_1^2-\jo{k_3} P_\text{I},
\end{align}
where $\jo{k_3}>0$ is a free parameter. Then, for any initial condition $(\hat{P}(t_0),P_\text{I}(t_0))$, \jo{we have that $\lim \hat{P}(t)\rightarrow P$ at exponential rate}.
\end{prop}
\begin{pf}
Define the error estimate
\begin{equation}\label{eq: error estimate P}
\tilde{P}(t)=\hat{P}(t)-P,
\end{equation}
then, along trajectories of \eqref{eq: dyn net + SD} it holds that
\begin{align}\label{eq: dynamics P tilde}
\dot{\tilde{P}}(t) & =\dot{\hat{P}}(t)=-\jo{k_3} v_1{C_1}\dot{v}_1+\dot{P}_\text{I}(t) \nonumber\\
& = -\jo{k_3}\left({v_1}i_1-P-v_1i_2 \right)+\dot{P}_\text{I}(t) \nonumber \\
& = -\jo{k_3} \tilde{P}(t) .
\end{align}
This implies that for any initial condition $\tilde{P}(t_0)$, $\tilde{P}(t)$ is given by
\begin{equation*}
\tilde{P}(t)=e^{-\jo{k_3}(t-t_0)}\tilde{P}(t_0).
\end{equation*}
Hence, $\tilde{P}(t)$ converges to zero at exponential rate. \qed
\end{pf}
\begin{rem}
The static controller defined in \eqref{eq: u simpler } depends on the components $\bar{x}_1$, $\bar{x}_2$, and $\bar{x}_3$ of the desired steady state value $\bar{x}\in\mathcal{E}$ (see \jo{Proposition \ref{prop: assign equil}}). In particular, $\bar{x}_2$ is assumed to be specified exactly. However, $\bar{x}_3$ depends linearly on the now assumed unknown CPL power $P$. In the next proposition we show that the controller is able to achieve the stabilization of $\bar{x}$ even when the designed on-line estimate for $P$ is being used. 
\end{rem}

\begin{prop}\label{prop: adaptive controller}
Let $\jo{k_3}>0$ be chosen arbitrarily. \jo{For the controller  \eqref{eq: u simpler }, define its adaptive version as}
\jo{\begin{equation}\label{eq: control law adaptive}
u=\hat{\gamma}(x)=\gamma(x)\vert_{P=\hat{P}(t)},
\end{equation}}
where $\gamma$ is \jo{given} in equation \eqref{eq: gamma} and $\hat{P}(t)$ is the online estimate of $P$, computed from \eqref{eq: estimate for P}. Then, $(x,\hat{P})=(\bar{x},P)$, with $\bar{x}\in\mathcal{E}$, is an asymptotically stable equilibrium point of the extended dynamics conformed by \eqref{eq: dyn net + SD}, \eqref{eq: estimate for P} and \eqref{eq: control law adaptive}.
\end{prop}
\begin{pf}
First, let us note that there exist suitable continuous mappings $\lambda:x\mapsto \lambda(x)$, $\mu:\mapsto \mu(x)$, and $\xi:x\mapsto \xi(x)$, not written here for space reasons,
such that the control laws $\gamma$, defined in \eqref{eq: u simpler }, can be written as
\begin{equation}\label{eq: controller dependent of P}
\gamma(x)=\lambda(x) P^2+\mu(x)P+\xi(x).
\end{equation}
If we substitute $P$ by its estimate $\hat{P}(t)$ given in \eqref{eq: estimate for P}, which in turn can be written in terms of the error estimate $\tilde{P}(t)$ defined in \eqref{eq: error estimate P}, then we can write an estimate of $\gamma$, denoted by $\hat{\gamma}$, as follows
\begin{align}\label{eq: control law using power estimate}
\hat{\gamma}(x) &=\lambda(x)\hat{P}^2(t)+\mu(x)\hat{P}(t)+\xi(x) \nonumber\\
& = \lambda(x)\left(P+\tilde{P}(t) \right)^2+\mu(x)\left(P+\tilde{P}(t) \right)+\xi(x) \nonumber \\
& =\gamma(x)+\epsilon\left(x,\tilde{P}(t)\right),
\end{align}
where
\begin{equation*}
\epsilon\left(x,\tilde{P}(t)\right)=\tilde{P}(t)\left[\lambda(x)\left(2P+\tilde{P}(t) \right)+\mu(x) \right] .
\end{equation*}
Notice that we recover the controller \eqref{eq: controller dependent of P}, which assumes an exact knowledge of $P$, plus a vanishing term. Now, we explore the stability of the overall system conformed by \eqref{eq: dyn net + SD}, \eqref{eq: dynamics P tilde} and \eqref{eq: control law using power estimate}. Let $\bar{x}\in\mathcal{E}$ a desired equilibrium point to be stabilized. Using $u=\hat{\gamma}(x)$ as a control law, the system \eqref{eq: dyn net + SD} adopts the form
\begin{align}\label{eq: closed loop P estimate}
\dot{x}& =f(x)+g(x)\cdot\hat{\gamma}(x) \nonumber \\
& =f(x)+g(x)\cdot\left(\gamma(x)+\epsilon\left(x,\tilde{P}(t) \right) \right) \nonumber\\
& = f(x)+g(x)\cdot\gamma(x)+g(x)\cdot \epsilon\left(x,\tilde{P}(t) \right).
\end{align}
Observe that this system is in cascade with \eqref{eq: dynamics P tilde}. Furthermore, if $\tilde{P}(t)$ is assumed to be zero, then $\bar{x}$ is an asymptotically stable equilibrium point, see Proposition \ref{prop: stabilization result ideal P}. Recalling that $\tilde{P}=0$ is an exponentially stable equilibrium point for \eqref{eq: dynamics P tilde},  we use \cite[Proposition 4.1]{sepulchre2012constructive}, to conclude that $(x,\tilde{P})=(\bar{x},0)$ is an asymptotically equilibrium point of the cascaded system conformed by \eqref{eq: dynamics P tilde} and \eqref{eq: closed loop P estimate}. \qed
\end{pf}


\section{Numerical validation}\label{sec: numerical validation}

In this section, we numerically evaluate the performance of the proposed adaptive controller.  The system physical parameters are taken according to Table \ref{tab:parameters}, \jo{and the parameters for the adaptive controller, presented in equation \eqref{eq: control law adaptive}, are taken as}
\begin{equation*}
k_1=30,~~k_2=0.78,~~k_3=1000.
\end{equation*}
{\jo{Simulations for the closed-loop system \eqref{eq: dyn net + SD} and \eqref{eq: control law adaptive} were done taking the initial condition}
\begin{align*}
x(0) & =\text{col}\left(40, ~12, ~31.6667, ~612.3611\right),
\end{align*}
which corresponds to the equilibrium point  $\bar{x}\vert_{P=100}$, with the particular value of $\bar x_2$  proposed in equation \eqref{eq: desired bar x2 optimal P}, given by
\begin{equation*}
\jo{\bar x_2} =\frac{1}{2}E= 12~\text{V},
\end{equation*}
see Prop. \ref{prop: assign equil}. Then, at $t=1~\mu s$, a step change in the CPL power, from $P=100$ W to $P=479~\text{W}$ is introduced. The control objective is to stabilize the {\em new} equilibrium point
\begin{align*}
\bar{x}\vert_{P=479}& =\text{col}\left(40, ~12, ~0.0833, ~31.6222\right) \in \mathcal{E}.
\end{align*}
We point out that the controller is be able to keep the same desired value for $\bar x_2$ regardless of the change in the value of $P$. Nonetheless, the rest of the coordinates of $\bar x$ \jo{change}, see equation \eqref{eq: bar x assign equi}. For clarity we use a logarithmic scale for the time to better appreciate the transient response of the system.\

Fig. \ref{fig:simulation_1a} shows the plot of $\jo{x_2(t)}$. Clearly, the controller is able to maintain the value of $x_2$ around the desired value of $\bar x_2=\frac{1}{2}E $, in spite of the change in the power consumed by the CPL.\

Fig. \ref{fig:simulation_1b} shows \jo{the plot of $u(t)$}. Observe that its value is always bounded within zero and one. Hence, the controller is physically realizable with the proposed DC-DC power converter.

{We underscore that the new CPL power satisfies the upper bound for stability established in \eqref{eq: new upper bound for P}, which with our numerical parameters reads as
\begin{equation*}
P<479.85~\text{W}.
\end{equation*}
This condition clearly relaxes the {\em necessary} condition for stability presented in Proposition \ref{prop: stab one port open loop}, which says that if the CPL power satisfies
\begin{equation*}
P> \frac{{E}^2 C_1 L_1 r_1}{\left(L_1+C_1 r_1^2\right){}^2}=276.9~\text{W},
\end{equation*}
then, the network {\em without} the shunt damper is unstable. Observe that the new CPL power value is very close to the maximum admissible value for existence of equilibria written in \eqref{eq: exis steady states power}, which in our case corresponds to $P\leq 480~\text{W}$. Hence, the addition of the shunt damper, operating under our adaptive controller, achieves  the stabilization of a desired equilibrium point $\bar x \in \mathcal{E}$ for a wide range of values \jo{for the CPL}.} \ju{For the sake of completeness, in Fig. \ref{fig:simulation_3a}, we present the plots of the state variables $x_1(t)$, $x_3(t)$, $x_4(t)$ and the estimated power against the time.}.\

In Fig. \ref{fig:simulation_2a} we present the plot of $x_2(t)$ for two scenarios: with the shunt damper and without it. This results were obtained by repeating the previous simulation but now taking the final value of $P$ as $P=260~\text{W}$ at $t=1~\text{ms}$. For network without the shunt damper, the steady state voltage $\bar x_2$ clearly decreases with the increase in the value of the CPL power. Additionally, an oscillatory behavior is present during the transient. Lastly, when the shunt damper is connected, the stabilization of $x_2$ around a desired value is achieved, independently of the CPL consumption.\

\begin{figure}
	\centering
		\includegraphics[width=\linewidth]{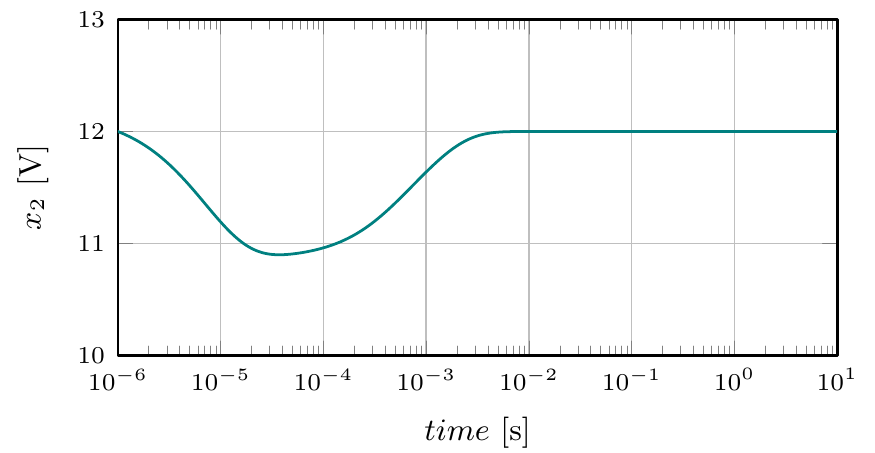}
	\caption{Plot of $x_2(t)$ for the system \eqref{eq: dyn net + SD} using the adaptive controller of equation \eqref{eq: control law adaptive}. At $t=1~\mu \text{s}$ a step change from $P=100$ W to $P=479$ W is introduced.}
	\label{fig:simulation_1a}
\end{figure}

\begin{figure}
	\centering
	\includegraphics[width=\linewidth]{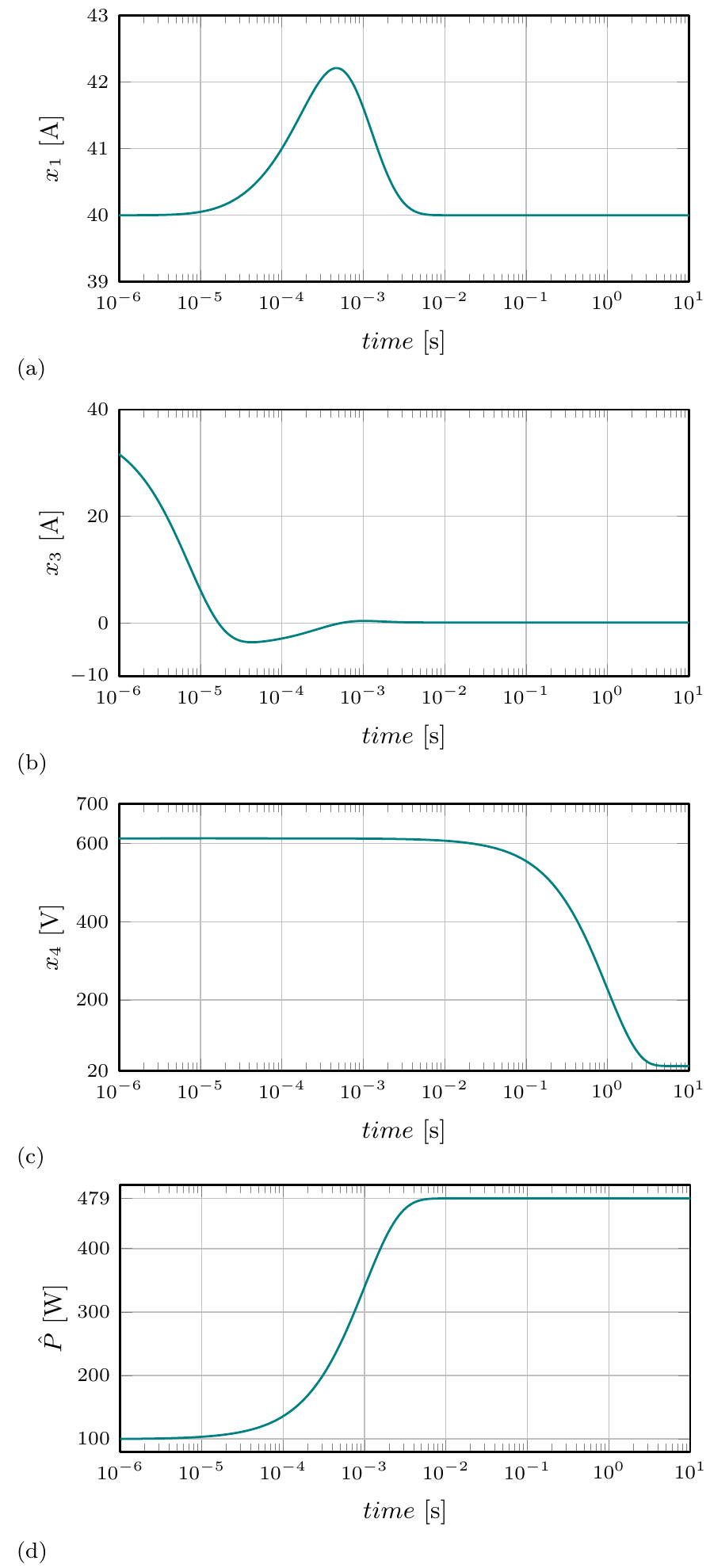}
	\caption{Plot of (a) $x_1(t)$, (b) $x_3(t)$,  (c) $x_4(t)$ and (d) $\hat{P}(t)$  for the system \eqref{eq: dyn net + SD} using the adaptive controller of equation \eqref{eq: control law adaptive}. At $t=1~\mu \text{s}$ a step change from $P=100$ W to $P=479$ W is introduced.}
	\label{fig:simulation_3a}
\end{figure}

\begin{figure}
	\centering
		\includegraphics[width=\linewidth]{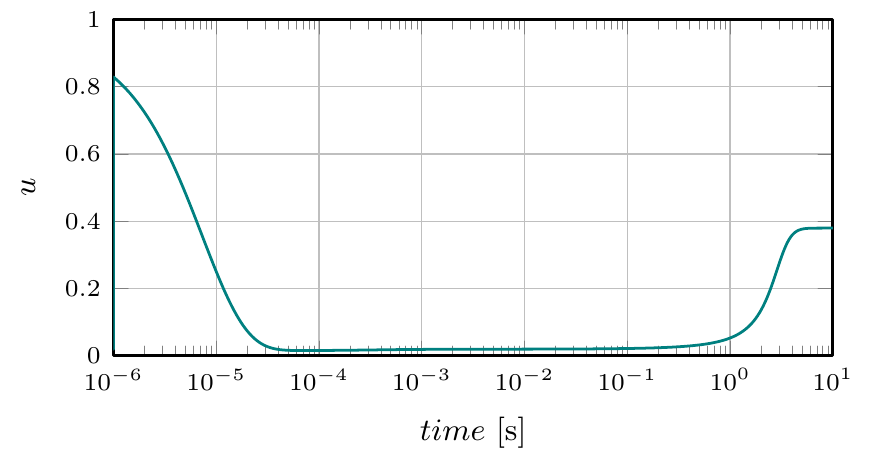}
	\caption{Plot of $u(t)$ for the system \eqref{eq: dyn net + SD} using the adaptive controller of equation \eqref{eq: control law adaptive}. At $t=1~\mu \text{s}$ a step change from $P=100$ W to $P=479$ W is introduced.}
	\label{fig:simulation_1b}
\end{figure}

\begin{figure}
	\centering
		\includegraphics[width=\linewidth]{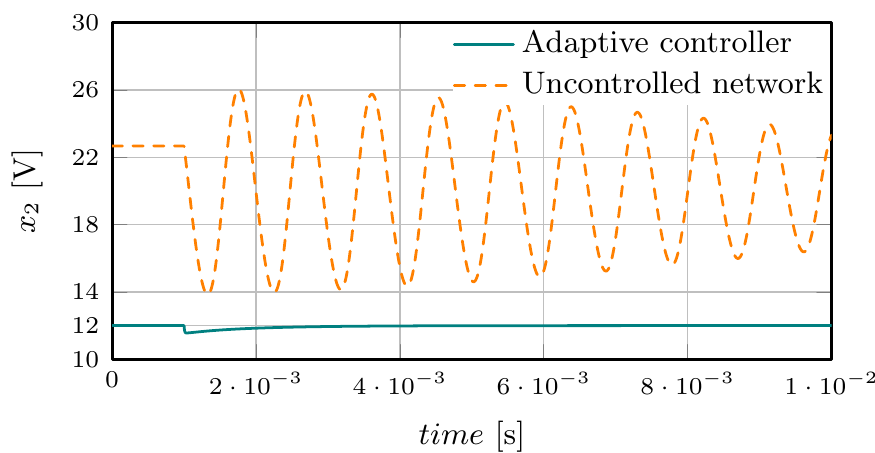}
	\caption{Plot of $x_2(t)$ with and without the shunt damper. At $t=1~\text{ms}$ a step change from $P=100$ W to $P=260$ W is introduced.}
	\label{fig:simulation_2a}
\end{figure}
\begin{table}[htb]
	\caption{Parameters for the circuit in Fig. \ref{fig: shunt damper}}
	\label{tab:parameters}
	\centering
	\begin{tabular}{c|c|c|c}
		\hline
		$r_1=0.3 \ \Omega$ & $L_1 = 85.0 \ \mu\mathrm{H}$ &  $C_1 = 200 \ \mu\mathrm{F}$ & $E = 24.0 \ \mathrm{V}$ \\ \hline
		$r_2=5 \ m\Omega$ & $L_2 = 100 \ \mu\mathrm{H}$ &  $C_2 = 1.0 \ \mathrm{mF}$ & $r_3=1 \ k\Omega$ \\
		\hline
	\end{tabular}
\end{table}
%
%
%
\ju{
	\begin{rem}\label{rem: energetic effic}
		In the simulations, we have focused on the stabilization of $\bar x$ with a fixed $\bar x_2$ as proposed in Corollary \ref{cor: choice of bar x2}. However, other values for $\bar x_2$ can be chosen as long as inequalities \eqref{eq: bounds for P in terms of x2bar} are satisfied. Diverse numerical experiments showed a dependency between the chosen value of $\bar x_2$ and the energetic efficiency of the power converter. Further analysis of this phenomenon is left as a future research.
	\end{rem}
}
\section{Conclusions and future work}\label{sec: conclusions and future work}
In this paper we have proposed a nonlinear stabilization method for a DC small-scale power system supplying electric energy to a CPL. This is done by incorporating an active shunt damper, consisting \jo{of} a controlled DC-DC power converter connected at the point of common coupling, exactly between the feeder and the load. \jo{Using s-PBC and I\&I theories, a nonlinear adaptive control law for the shunt damper is designed.}. It permits the stable operation of the network for a wide range of values of the CPL and is able to relax some necessary stability bounds that are imposed when the system operates {\em without} the shunt damper. Through realistic numerical simulations, we have \jo{illustrated} the satisfactory behavior of the designed controller.\

The results of this paper can be extended in the following directions:\

- Explicitly compute estimates for the Region of Attraction of the system in closed-loop. Particularly, for the case when the power estimate, designed in Section \ref{sec: power estimator and stab analysis}, is used.\

- Theoretically evaluate the robustness, against parameter uncertainty, of the proposed adaptive control.\

- Design \jo{an observer} for the variable $x_1$, which is, in some practical scenarios, difficult to measure. We underscore that both, our controller and our power estimator, explicitly depend on this value.\

\ju{
	-Further investigate the dependency between the chosen value $\bar x_2$ and the energetic efficiency of the power converter, see Remark \ref{rem: energetic effic}.\
}

- Analyze the viability of applying the present stabilization result to the case of multi-port networks with a distributed array of CPLs.

%

\bibliography{ref}

\end{document}